\documentclass[12pt]{article}
\usepackage{amsmath,amsfonts,amssymb,amsthm,graphicx,cite}
\usepackage{mathrsfs} 

\newtheorem{theorem}{Theorem}[section]
\newtheorem{lemma}[theorem]{Lemma}
\newtheorem{prop}[theorem]{Proposition}
\newtheorem{cor}[theorem]{Corollary}
\newtheorem{remark}[theorem]{Remark}

\newcommand{\Ref}[1]{(\ref{#1})}

\newcommand{\QED}{\hfill$\square$}

\newcommand{\ee}[1]{{\rm e}^{#1}}
\newcommand{\ii}{{\rm i}}

\newcommand{\tet}{\theta}

\newcommand{\vx}{{\bf x}}
\newcommand{\vX}{{\bf X}}
\newcommand{\vy}{{\bf y}}
\newcommand{\vz}{{\bf z}}
\newcommand{\vm}{{\bf m}}

\newcommand{\R}{{\mathbb R}}
\newcommand{\C}{{\mathbb C}}
\newcommand{\Z}{{\mathbb Z}}
\newcommand{\N}{{\mathbb N}}

\newcommand{\cW}{{\mathcal  W}}
\newcommand{\cH}{{\mathcal  H}}
\newcommand{\cE}{{\mathcal  E}}
\newcommand{\cP}{{\mathcal  P}}
\newcommand{\cN}{{\mathcal  N}}

\begin{document}

\begin{flushright}
July 12, 2010
\end{flushright}
\vspace{.4cm}

\begin{center}

{\Large \bf Source identity and kernel functions for elliptic Calogero-Sutherland type systems} 

\vspace{1 cm}

{\large Edwin Langmann}\footnote{langmann@kth.se}\\
\vspace{0.3 cm} 

\textit{Theoretical Physics, AlbaNova, SE-106 91 Stockholm, Sweden}

\end{center}

\begin{abstract}
Kernel functions related to quantum many-body systems of Calogero-Suther\-land type are discussed, in particular for the elliptic case.  The main result is an elliptic generalization of an identity due to Sen that is a source for many such kernel functions. Applications are given, including simple exact eigenfunctions and corresponding eigenvalues of Chalykh-Feigin-Veselov-Sergeev-type deformations of the elliptic Calogero-Sutherland model for special parameter values.

\medskip

\noindent {MSC-class: 81Q05, 16R60}
\end{abstract}

\section{Introduction}
\label{sec1} 
Sen proved in \cite{Sen} that the quantum many body Hamiltonian 
\begin{equation}
\label{cH} 
\cH := -\sum_{J=1}^{\cN} \frac1{m_J}\frac{\partial^2}{\partial X_J^2} +
\sum_{1\leq J<K\leq \cN}\gamma_{JK} V(X_J-X_K) 
\end{equation} 
for $V(r)=1/[4\sin^2(r/2)]$ and the coupling constants 
\begin{equation}
\label{gJK} 
\gamma_{JK} :=  \lambda(m_J+m_K)(\lambda m_Jm_K-1)
\end{equation} 
has the following exact groundstate
\begin{equation}
\label{Phi0} 
\Phi_0(\vX) := \prod_{1\leq J<K\leq \cN} \theta(X_J-X_K)^{\lambda m_Jm_K}
\end{equation}
with $\theta(r)=\sin(r/2)$, for arbitrary particle number $\cN$, coupling parameter $\lambda>0$, and particle masses $m_J/2>0$. He was also able to compute the corresponding ground state energy $\cE_0$ exactly \cite{Sen}. Sen's result is a  generalization of a well-known result for the Calogero-Sutherland (CS)  \cite{C,Su} model (corresponding to the special cases where the particles are non-distinguishable and $m_J=1$ for all $J$). As discussed in \cite{HL}, Sen's identity $(\cH-\cE_0)\Phi_0(\vX)=0$ holds true for arbitrary real parameters $m_J$ and $\lambda$, and this (obvious) generalization is interesting since it has many special cases that provide means to compute exact eigenfunctions and corresponding eigenvalues of the CS model and, more generally, of Chalykh-Feigin-Veselov-Sergeev- (CFSV-) type deformations \cite{CFV,Sergeev,SV} of the CS model. The latter are given by differential operators
\begin{equation}
\label{HNtN} 
\begin{split} 
H_{N,\tilde{N}}(\vx,\tilde{\vx}):=-\sum_{j=1}^N\frac{\partial^2}{\partial x_j^2} + 
\sum_{1\leq j<k\leq N} 2\lambda(\lambda-1)V(x_j-x_k) \\
+\lambda\sum_{J=1}^{\tilde{N}}\frac{\partial^2}{\partial \tilde{x}_J^2} + 
\sum_{1\leq J<K\leq \tilde{N}} \frac{2(\lambda-1)}{\lambda}V(\tilde{x}_J-\tilde{x}_K) \\
+\sum_{j=1}^N \sum_{K=1}^{\tilde{N}} 2(1-\lambda)V(x_j-\tilde{x}_K)
\end{split} 
\end{equation}    
and reduce to the CS Hamiltonian for $\tilde{N}=0$ ($N$ and $\tilde{N}$ are non-negative integers such that $N+\tilde{N}\geq 1$, and we write $\vx$ short for $(x_1,\ldots,x_N)$ and similarly for $\tilde{\vx}$). 
The above mentioned special cases are identities of the form
\begin{equation}
\label{kernelfunction} 
\Bigl(  H_{N,\tilde{N}}(\vx,\tilde{\vx}) - H_{M,\tilde{M}}(\vy,\tilde{\vy})-C  \Bigr) F(\vx,\tilde{\vx},\vy,\tilde{\vy})=0 
\end{equation} 
for some function $F$ and some constant $C$. We refer to such $F$ as {\em kernel function} of the pair of differential operators $(H_{N,\tilde{N}}(\vx,\tilde{\vx}), H_{M,\tilde{M}}(\vy,\tilde{\vy}))$ in the following.

In this paper we give a generalization of Sen's identity to the elliptic case, i.e.\ to the case where $V(r)$ is (essentially) the Weierstrass elliptic function $\wp(r)$. We also point out various special cases of this elliptic identity that provide means to compute exact eigenfunctions and eigenvalues of the CFSV-deformed elliptic CS (eCS) differential operator in \Ref{HNtN}, in generalization of results in  \cite{HL} (trigonometric limit) and \cite{EL0,EL1,EL2,EL3} (eCS model).

Our Corollary~\ref{cor1} generalizes and unifies identities given in \cite{EL4}, (6)--(12) (the latter correspond to the special cases $\tilde{N}=\tilde{M}=0$ and $\tilde{N}=M=0$). Generalizations of the identities in \cite{EL4} to the relativistic (Ruijsenaars) generalization of the CS-model \cite{R} with a restriction on parameters as in \Ref{specialcase} were recently given by Komori, Noumi, and Shiraishi \cite{KNS}; see also \cite{R1} and references therein for related results. Identities like in \Ref{kernelfunction} were also used in other works to construct eigenfunctions CS-type operators, including \cite{KNS,KMS,FV}. 

We find in this paper exact eigenfunctions of elliptic CS-type differential operators that can be represented by simple, explicit formulas (Corollaries~\ref{cor2} and \ref{cor3}). The only other similar result in the literature we are aware of are $BC_N$ variants of the eCS model \cite{OP} (also known as Inozemtsev model \cite{I}) that were found by G\'omez-Ullate, Gonz\'alez-L\'opez, and Rodr\'iguez to be quasi-exactly solvable for certain special parameter values \cite{GGR}; see also \cite{T,FGGRZ}. 

The results in the present paper and in \cite{KNS} suggest that there should exist a generalization of Sen's identity to the relativistic case, and this identity might be easier to prove than its special cases found in \cite{KNS}. This result would also be interesting since, as we expect, it should provide means to obtain results for CFSV-type deformations of the Ruijsenaars models. We also conjecture that there exists a Sen-like identity for Inozemtsev-type models. 

The rest of this paper is organized as follows. In Section~\ref{sec2} we present our results, and in Section~\ref{sec3} we discuss applications. Proofs can be found in Section~\ref{sec4}. Some results on elliptic functions that we need are collected in an appendix. 

\section{Results}
\label{sec2} 
We find it convenient to fix the elliptic function periods as $2\omega_1=2\pi$ and $2\omega_2=\ii\beta$ with $\beta>0$, and we add a constant to $\wp(r)$ so that the trigonometric limit $\beta\to \infty$ becomes simple:\footnote{Our conventions for special functions are as in \cite{AS}. Details can be found in \cite{EL2}, Appendix~A.}
\begin{equation}
\label{V}
V(r) : = \sum_{m\in\Z}\frac1{4\sin^2[(r+\ii\beta m)/2]} = 
\wp(r|\pi,\ii\beta/2) +c_0
\end{equation} 
with
\begin{equation}
\label{c0c1} 
c_0:= \frac{\eta_1}{\omega_1}= c_1-\sum_{m=1}^\infty \frac1{2\sinh^2(\beta m/2)},\quad c_1:=\frac1{12} . 
\end{equation} 
We also need
\begin{equation}
\label{tet} 
\tet(r) := \sin(r/2)\prod_{m=1}^\infty(1-2q^{2m}\cos(r)+q^{4m}) ,\quad q:=\ee{-\beta/2} 
\end{equation} 
proportional to the Jacobi theta function $\vartheta_1(r/2,q)$. We use the short hand notation 
\begin{equation}
\label{vmn} 
|\vm^n|:= \sum_{J=1}^{\cN} m_J^n, \quad n=1,2,3.  
\end{equation} 

Our main result is the following. 

\begin{prop}
\label{prop1} 
Let $\cN\in\N$, and $\lambda$ and $m_J$ complex and non-zero for $J=1,2,\ldots,\cN$. Then the differential operator in \Ref{cH}--\Ref{gJK} with $V(r)$ in \Ref{V} and the function $\Phi_0(\vX)$ in \Ref{Phi0} with $\tet(r)$ in \Ref{tet} obey the identity
\begin{equation}
\label{Id1} 
\left(\cH + 2\lambda|\vm|\frac{\partial}{\partial\beta} -\cE_0\right)\Phi_0(\vX) =0  
\end{equation} 
with the constant
\begin{equation}
\label{cE0} 
\cE_0 = \lambda^2\Bigl( (|\vm^2||\vm|-|\vm^3|)c_0 +(|\vm|^2-|\vm^2|)|\vm|c_1\Bigr) 
\end{equation}
and $c_{0,1}$ in \Ref{c0c1}. 
\end{prop}

(Proof in Section~\ref{sec4.1}.) 

In the following we are mainly interested in cases where $\lambda>0$ and all $m_J$ are real. 

Due to the $\beta$-derivative term in \Ref{Id1} it is not possible to interpret $\Phi_0(\vX)$ as eigenfunction of $\cH$ in general. However, regarding $\tau:=\beta/(2|\vm|\lambda)$ as (imaginary) time, one can interpret \Ref{Id1} as (Wick rotated) time evolution equation for quantum many-body systems with peculiar time dependent two-body potentials, and Proposition~\ref{prop1} gives an explicit solution of this equation. It could be interested to explore this interpretation of \Ref{Id1} and its consequences further. However, in the rest of this paper we only discuss a very different application: one can use \Ref{Id1} as a source to obtain kernel functions for pairs of eCS-type differential operators as in \Ref{HNtN} \cite{HL}. 

The key observation is this: One can choose the parameters $\cN$ and $m_J$ such that $\cH$ in \Ref{cH} is (essentially) a difference of two operators $H_{N,\tilde{N}}(\vx,\tilde{\vx})$ and $H_{M,\tilde{M}}(\vy,\tilde{\vy})$ \cite{HL}. Indeed, $\gamma_{JK}=0$ in the following four cases: (i) $m_J=1$ and $m_K=-1$, (ii) $m_J=1$ and $m_K=1/\lambda$, (iii) $m_J=-1/\lambda$ and $m_K=-1$, and (iv) $m_J=-1/\lambda$ and $m_K=1/\lambda$. One thus can divide the variables $\vX$ in four groups $\vx$, $\tilde{\vx}$, $\vy$ and $\tilde{\vy}$ where the ``mass parameter'' $m_J$ is the same for all variables in each group. Choosing the $m_J$ as $1$, $-1/\lambda$, $-1$, and $1/\lambda$ in these four groups the first two groups of variables decouple from the last two groups, and Proposition~\ref{prop1} implies the following.

\begin{cor}
\label{cor1}
Let $N$, $\tilde{N}$, $M$, and $\tilde{M}$ be non-negative integers and $\lambda$ complex and non-zero. Then the function
\begin{equation}
\label{FNtNMtM} 
\begin{split}
F_{N,\tilde{N},M,\tilde{M}}(\mathbf{x},\mathbf{\tilde{x}}, \mathbf{y},\mathbf{\tilde{y}}) :=  &\Psi_0^{N,\tilde{N}}(\vx,\tilde{\vx})  \Psi_0^{M,\tilde{M}}(\vy,\tilde{\vy})\\ &\times \frac{\Bigl(\prod_{j=1}^N\prod_{J=1}^{\tilde{M}}\tet(x_j-\tilde{y}_K)\Bigr)\Bigl(\prod_{J=1}^{\tilde{N}}\prod_{k=1}^M \tet(\tilde{x}_J-y_k)\Bigr)}{\Bigl(\prod_{j=1}^N\prod_{k=1}^M \tet(x_j-y_k)^\lambda\Bigr)\Bigl( \prod_{J=1}^{\tilde{N}}\prod_{K=1}^{\tilde{M}} \tet(\tilde{x}_J-\tilde{y}_K)^{1/\lambda}\Bigr)}\\
\end{split}
\end{equation} 
with
\begin{equation}
\label{PsiNtN} 
 \Psi_0^{N,\tilde{N}}(\vx,\tilde{\vx}) := 
 \frac{\Bigl(\prod_{1\leq j<k\leq N}\tet(x_j-x_k)^{\lambda}\Bigr)\Bigl(\prod_{1\leq J<K\leq \tilde{N}}\tet(\tilde{x}_J-\tilde{x}_K)^{1/\lambda}\Bigr)}{\prod_{j=1}^N\prod_{J=1}^{\tilde{N}}\tet(x_j-\tilde{x}_J)}
\end{equation} 
and the differential operators $H_{N,\tilde{N}}(\vx,\tilde{\vx})$ and $H_{M,\tilde{M}}(\vy,\tilde{\vy})$ in \Ref{HNtN} obey the identity
\begin{equation}
\label{Id2} 
\Bigl(H_{N,\tilde{N}}(\vx,\tilde{\vx})-H_{M,\tilde{M}}(\vy,\tilde{\vy})+2[(N-M)\lambda-\tilde{N}+\tilde{M}]\frac{\partial}{\partial\beta} -C_{N,M,\tilde{N},\tilde{M}}\Bigr)F_{N,\tilde{N},M,\tilde{M}}(\mathbf{x},\mathbf{\tilde{x}}, \mathbf{y},
  \mathbf{\tilde{y}}) = 0 
\end{equation}
with the constant 
\begin{multline}
\label{CNtNMtM} 
C_{N,M,\tilde{N},\tilde{M}} = \Bigl( [N(N-1)-M(M-1)]\lambda^2 -(N+M)(\tilde{N}-\tilde{M})\lambda +(N-M)(\tilde{N}+\tilde{M})\\ -[\tilde{N}(\tilde{N}-1)-\tilde{M}(\tilde{M}-1)]/\lambda\Bigr)c_0 +\Bigl( (N-M)[(N-M)^2-N-M]\lambda^2\\ -[3(N-M)^2-N-M](\tilde{N}-\tilde{M})\lambda+(N-M)[3(\tilde{N}-\tilde{M})^2-\tilde{N}-\tilde{M}]\\-(\tilde{N}-\tilde{M})[(\tilde{N}-\tilde{M})^2-\tilde{N}-\tilde{M} ]/\lambda\Bigr)c_1
\end{multline}
and $c_{0,1}$ in \Ref{c0c1}. 
\end{cor} 

(Proof in Section~\ref{sec4.2}.) 

In cases where there is no $\beta$-derivative terms, i.e.\ if the parameters are such that 
\begin{equation}
\label{specialcase}
(N-M)\lambda=\tilde{N}-\tilde{M}
\end{equation} 
holds true, we obtain an identity as in \Ref{kernelfunction}, i.e.\ $F_{N,\tilde{N},M,\tilde{M}}$ in \Ref{FNtNMtM} is a kernel function for the pair of differential operators $(H_{N,\tilde{N}}(\vx,\tilde{\vx}),H_{M,\tilde{M}}(\vy,\tilde{\vy}))$ if \Ref{specialcase} holds true. Note that the restriction in \Ref{specialcase} is not present in the trigonometric limit \cite{HL}. 

\begin{remark}
\label{rem1} 
One can show that by redefining the elliptic functions
\begin{equation}
V(r)\to \wp(r|\pi,\ii\beta),\quad \tet(r)\to \vartheta(r/2,q)  
\end{equation} 
the identities in Proposition~\ref{prop1} and Corollary~\ref{cor1} hold true as they stand but with the constants redefined as follows
\begin{equation}
\begin{split}
\cE_0 &\to \lambda(\cN-1)|\vm|c_0\\
C_{N,\tilde{N},M,\tilde{M}}  &\to (N+\tilde{N}+M+\tilde{M}-1)[\lambda(N-M)-\tilde{N}+\tilde{M}]c_0 
\end{split} 
\end{equation}
(see Section~\ref{sec4.3} for details). This shows that choosing standard elliptic functions makes the constants in our result significantly simpler. Our choice has the advantage that the trigonometric limit $\beta\to \infty$ is obvious, whereas standard elliptic function requires a non-trivial multiplicative renormalization. Moreover, for the cases of interest to us the constants are simple anyway; see the remark below. 
\end{remark}

\begin{remark}
\label{rem2}
We are mainly interested in the special cases of \Ref{Id1} where $|\vm|=0$. In such a case the constants in \Ref{cE0} and \Ref{CNtNMtM} simplify significantly as follows,
\begin{equation}
\label{simple}
\begin{split}
\cE_0 &= -\lambda^2|\vm^3|c_0\\
C_{N,\tilde{N},M,\tilde{M}} &= [-\lambda^2(N-M)+(\tilde{N}-\tilde{M})/\lambda]c_0.
\end{split} 
\end{equation} 
\end{remark} 

\begin{remark}
\label{rem3}
At first sight it seems one could obtain a generalization of the identity in \Ref{Id2} from Proposition~\ref{prop1} by choosing the ``mass parameters'' for the four groups of variables as $m$, $-1/(m\lambda)$, $-m$, $1/(m\lambda)$ with arbitrary real $m\neq 0$. However, multiplying the identity thus obtained by $m$ and changing $\lambda$ to $\lambda/m^2$ one recovers \Ref{Id2}. We thus set $m=1$ without loss of generality. 
\end{remark}

\begin{remark}
\label{rem4}  
Replacing the set of parameters $(N,\tilde{N},M,\tilde{M},\lambda)$ by $(\tilde{N},N,\tilde{M},M,1/\lambda)$ leaves the identity in \Ref{Id2} invariant. This suggests that the duality transformation $(N,\tilde{N},\lambda)\to (\tilde{N},N,1/\lambda)$ should be an interesting symmetry of the differential operators in \Ref{HNtN}. This symmetry is a generalization of the well-known duality of the Jack polynomials (see e.g.\ \cite{McD}) corresponding to the special cases $\beta\to\infty$ and $\tilde{N}=0$; see \cite{HL} for further details and results in the trigonometric limit. 
\end{remark} 

In applications it is convenient to use the following slight generalization of the result in Corollary~\ref{cor1}. 

\begin{lemma}
\label{lem1} 
\noindent The identity in \Ref{Id2} remains true as it stands if one replaces
\begin{equation}
\label{subs2}
\begin{split} 
F_{N,\tilde{N},M,\tilde{M}}(\mathbf{x},\mathbf{\tilde{x}}, \mathbf{y},\mathbf{\tilde{y}})&\to c F_{N,\tilde{N},M,\tilde{M}}(\mathbf{x},\mathbf{\tilde{x}}, \mathbf{y},\mathbf{\tilde{y}})\ee{\ii v[|\vx|-|\vy|-(|\tilde{\vx}|-|\tilde{\vy}|)/\lambda]}\\ 
C_{N,M,\tilde{N},\tilde{M}} &\to C_{N,M,\tilde{N},\tilde{M}} + [N-M-(\tilde{N}-\tilde{M})/\lambda] v^2
\end{split}
\end{equation}
with arbitrary constants $v\in\R$ and $c\in\C\setminus\{0\}$, with $|\vx|:=\sum_{j=1}^N x_j$ etc.  
\end{lemma}

(Proof and physical interpretation in Section~\ref{sec4.4}.) 

As will become clear in the next section, this result allows one to remove exponential factors, corresponding to trivial center-of-mass contributions, from eigenfunctions. 

\section{Applications} 
\label{sec3} 
In this section we point out various interesting special cases of our results in the previous section. 

As already mentioned, it is not possible in general to interpret \Ref{Id1} as an eigenvalue equation.  However, in case the parameters are such that $|\vm|=0$ this is possible. One such case of interest to us is obtained from Corollary~\ref{cor1} setting $M=\tilde{M}=0$ and denoting $C_{N,\tilde{N},0,0}$ as $E_0$:

\begin{cor}
\label{cor2} 
Let $N$ and $\tilde{N}$ be non-negative integers and 
\begin{equation}
\label{cond1} 
\lambda=\tilde{N}/N. 
\end{equation} 
Then $\Psi_0^{N,\tilde{N}}(\mathbf{x},\mathbf{\tilde{x}})$ in \Ref{PsiNtN}  with $\tet(r)$ in \Ref{tet} is an exact eigenfunction of the differential operator $H_{N,\tilde{N}}(\mathbf{x},\mathbf{\tilde{x}})$ in \Ref{HNtN} with $V(r)$ in \Ref{V}, and the corresponding eigenvalue is
\begin{equation}
\label{E0}
\begin{split} 
E_0 = (N-\tilde{N}^2/N)c_0
\end{split} 
\end{equation} 
with $c_0$ in \Ref{c0c1}. 
\end{cor}

Another identity where one can construct simple eigenfunctions of the differential operator $H_{N,\tilde{N}}(\mathbf{x},\mathbf{\tilde{x}})$ is \Ref{Id2} for $M=1$, $\tilde{M}=0$, and $(N-1)\lambda-\tilde{N}=0$, i.e.\
 \begin{equation}
 \label{Id3} 
 \Bigl(H_{N,\tilde{N}}(\mathbf{x},\mathbf{\tilde{x}}) +  \frac{\partial^2}{\partial y^2} - C_{N,\tilde{N},1,0} \Bigr) c \Psi_0^{N,\tilde{N}}(\mathbf{x},\mathbf{\tilde{x}})\cP(\mathbf{x},\mathbf{\tilde{x}},y)\ee{\ii v(|\vx|-y-|\tilde{\vx}|/\lambda)} 
 \end{equation} 
with $C_{N,\tilde{N},1,0}=[-\lambda^2(N-1)-\tilde{N}/\lambda]c_0=[N-1-\tilde{N}^2/(N-1)]c_0$ and 
\begin{equation}
\label{cP} 
\cP(\mathbf{x},\mathbf{\tilde{x}},y) = \Bigl( \prod_{j=1}^N\tet(x_j-y)^{-\lambda}\Bigr)\Bigl(\prod_{J=1}^{\tilde{N}}\tet(\tilde{x}_J-y)\Bigr)  
\end{equation} 
with constants $v\in\R$ and $c\in\C\setminus\{0\}$ to be determined; we used the generalization of \Ref{Id2} pointed out in Lemma~\ref{lem1}. Inserting $\tet(x)=(\ii/2)\ee{-\ii x/2}\check{\tet}(\ee{\ii x})$ with
\begin{equation}
\label{ctet} 
\check\tet(z):=(1-z)\prod_{m=1}^\infty (1-q^{2m}z)(1-q^{2m}/z) 
\end{equation} 
analytical and non-zero for $q^2<|z|<1$, we observe that the function in \Ref{cP} is equal to $\check{\cP}(\vz,\tilde{\vz},\xi):= \prod_j\tet(z_j/\xi)^{-\lambda} \prod_J\tet(\tilde{z}_J/\xi)$ times $\exp(\ii[\lambda(|\vx|-y)-|\tilde{\vx}|]/2)$ with $\xi:=\ee{\ii y}$, up to some finite and non-zero multiplicative constant; here and in the following we use the notation
\begin{equation}
z_j:=\ee{\ii x_j},\quad \vz:=(z_1,\ldots,z_N) 
\end{equation} 
and similarly for $\tilde{\vx}$ (we inserted $N\lambda-\tilde{N}=\lambda$ to simplify the exponential factor). We thus find that, for $v=-\lambda/2$ and suitable $c\neq 0$, $c \cP(\mathbf{x},\mathbf{\tilde{x}},y)\exp{(\ii v[|\vx|-y-|\tilde{\vx}|/\lambda]}$ in \Ref{Id3} is identical with $\check{\cP}(\vz,\tilde{\vz},\xi)$. This latter function can be analytically continued to the annulus $1<|\xi|<1/q^2$ in the complex $\xi$-plane and is there equal to its Taylor series $\sum_{n\in\Z} \xi^{-n} \cP_n(\vz,\tilde{\vz})$ with functions $\cP_n(\vz,\tilde{\vz})$ that can be computed by contour integrals; see \Ref{cPn} below. Inserting this and $(\partial^2/\partial y^2)\xi^{-n} = -n^2 \xi^{-n}$ and comparing equal powers of $\xi$ we obtain the following. 

\begin{cor}
\label{cor3} 
Let $N\geq 2$ and $\tilde{N}\geq 1$ be integers and 
\begin{equation}
\lambda=\tilde{N}/(N-1). 
\end{equation} 
Then  the differential operator $H_{N,\tilde{N}}(\mathbf{x},\mathbf{\tilde{x}})$ in \Ref{HNtN} with $V(r)$ in \Ref{V} has the following exact eigenfunctions labeled by integers $n$, 
\begin{equation}
\Psi_{n}(\mathbf{x},\mathbf{\tilde{x}})= \Psi_0^{N,\tilde{N}}(\mathbf{x},\mathbf{\tilde{x}})\cP_n(\vz,\tilde{\vz}) 
\end{equation} 
with $\Psi_0^{N,\tilde{N}}(\mathbf{x},\mathbf{\tilde{x}})$ in \Ref{PsiNtN} and $\tet(r)$ in \Ref{tet}, and 
\begin{equation}
\label{cPn} 
\cP_n(\vz,\tilde{\vz})= 
\oint\frac{d\xi}{2\pi\ii \xi} \xi^n \Bigl( \prod_{j=1}^N\check\tet(\ee{\ii x_j}/\xi)^{-\lambda}\Bigr)\Bigl( \prod_{J=1}^{\tilde{N}}\check\tet(\ee{\ii \tilde{x}_J}/\xi)\Bigr) 
\end{equation}
with $\check\tet(z)$ in \Ref{ctet} and the integration contour a circle $|\xi|=R$ of radius $1<R<1/q^2$. Moreover, the corresponding eigenvalue is 
\begin{equation}
E(n) = n^2 + [N-1-\tilde{N}^2/(N-1)]c_0
\end{equation} 
with the constant $c_0$ in \Ref{c0c1}. 
\end{cor}

There are many generalizations of the results above that are, however, more complicated in general. A particularly interesting special case of the identity in \Ref{Id2} is for $\tilde{N}=N$ and $\tilde{M}=M$, i.e.\
\begin{equation}
\label{remarkable} 
\Bigl(H_{N,\tilde{N}}(\vx,\tilde{\vx})-H_{N,\tilde{N}}(\vy,\tilde{\vy})\Bigr)F_{N,\tilde{N}}(\mathbf{x},\mathbf{\tilde{x}}, \mathbf{y},  \mathbf{\tilde{y}}) = 0 
\end{equation} 
with $F_{N,\tilde{N}}:= F_{N,\tilde{N},N,\tilde{N}}$ in \Ref{FNtNMtM}. The latter generalizes a result in \cite{EL0} for the eCS model (special case $\tilde{N}=0$) that can be used to construct a perturbative solution of the eCS model to all orders \cite{EL2,EL3}. It should be straightforward to generalize this solution and construct eigenfunctions and corresponding eigenvalues of the CFSV-deformed eCS differential operators. Similarly as in \cite{HL}, one should also be able to construct different representations of these eigenfunctions and eigenvalues starting from any identity in \Ref{Id2} whenever \Ref{specialcase} holds true. The complexity of such a representation is determined by $M+\tilde{M}$: the smaller the latter the smaller the complexity \cite{HL}. The results in Corollaries~\ref{cor2} and \ref{cor3} correspond to cases with the smallest possible complexities $M+\tilde{M}=0$ and $1$, respectively. It would be interesting to study these solutions in more detail, but this is beyond the scope of the present paper. 
   
\section{Proofs}
\label{sec4} 
This section contains the proofs of the results stated in Section~\ref{sec2}. 
 
 \subsection{Proposition~\ref{prop1}}
 \label{sec4.1} 
The proof of Proposition~\ref{prop1} below is a straightforward computations using functional identities of elliptic functions collected in Appendix~\ref{appA}. 

With $\Phi_0$ in \Ref{Phi0}  we compute
\begin{equation}
\label{defW}
\cW := \frac1{\Phi_0} \sum_{J=1} \frac1{m_J} \frac{\partial^2}{\partial X_J^2} \Phi_0
\end{equation} 
where we suppress the common argument $\vX$ of the functions $\cW$ and $\Phi_0$, here and in the following. Straightforward computations give 
\begin{equation}
\cW = \sum_{J=1}^{\cN} \Bigl( \sum_{K\neq J} \lambda m_K\phi'(X_J-X_K)  +\sum_{K\neq J} \lambda m_Jm_K\phi(X_J-X_K) \sum_{L\neq J} \lambda m_L\phi(X_J-X_L) \Bigr) 
\end{equation} 
with $\phi(r)$ in \Ref{phi} and $\phi'(r):=\partial\phi(r)/\partial r$; we used \Ref{evenodd}. We write $\cW=\cW_1+\cW_2$ with
\begin{equation}
\label{W1} 
\cW_1 = \sum_{1\leq K<J\leq\cN} \Bigl( \lambda (m_J+m_K) \phi'(X_J-X_K) + \lambda^2 m_J m_K(m_J+m_K) \phi(x_J-x_K)^2  \Bigr)  
\end{equation} 
the sum of all two-body terms and
\begin{equation} 
\label{W2} 
\begin{split} 
\cW_2 =  \sum_{1\leq J<K<L\leq\cN} & \lambda^2 m_J m_K m_L \Bigl( \phi(X_J-X_K)\phi(X_J-X_L)\\ & +\phi(X_K-X_L)\phi(X_K-X_J)+\phi(X_L-X_J)\phi(X_L-X_K)\Bigr) 
\end{split} 
\end{equation} 
all three-body terms; all sums were made symmetric with respect to the summation indices using \Ref{evenodd}. Inserting the identities in \Ref{p} and \Ref{pp} in \Ref{W1} we obtain 
\begin{equation} 
\cW_1 = \sum_{1\leq J<K \leq\cN} \Bigl( \gamma_{JK}V(X_J-X_K) - \lambda^2m_Jm_K(m_J+m_K) [2f(x_J-x_K)+c_0 ]\Bigr) 
\end{equation} 
with $\gamma_{JK}$ in \Ref{gJK}. Using \Ref{ppp} and \Ref{evenodd} for $x=x_J-x_K$, $y=-(x_J-x_L)$ and $z=x_K-x_L$ we find
\begin{equation}
\begin{split} 
\cW_2 = & -\sum_{1\leq J<K<L\leq\cN} \lambda^2 m_J m_K m_L [ f(x_J-x_K) + f(x_J-x_L) + f(x_K-x_L)] \\  = & -2\sum_{1\leq J<K\leq \cN} \sum_{L\neq J,K} \lambda^2 m_J m_K m_L f(x_J-x_K) . 
\end{split} 
\end{equation} 
Adding $\cW_1$ and $\cW_2$ and inserting $\sum_{L\neq J,K} m_L = |\vm| - m_J - m_K$ gives 
\begin{equation} 
\label{Wb} 
\cW =  \sum_{1\leq J<K\leq\cN} \Bigl( \gamma_{JK}V(X_J-X_K)  -2\lambda^2 |\vm|m_Jm_Kf(x_J-x_K) -\lambda^2 c_0 m_Jm_K(m_J+m_K) \Bigr).  
\end{equation} 
From \Ref{Phi0} and \Ref{f} we conclude  
\begin{equation}
\frac1{\Phi_0}\frac{\partial}{\partial\beta}\Phi_0 = -\sum_{1\leq J<K\leq \cN}  \lambda m_Jm_K[ f(x_J-x_K) - c_1]  
\end{equation} 
and thus \Ref{defW} and \Ref{Wb} imply
\begin{equation}
\frac1{\Phi_0} \sum_{J=1} \frac1{m_J} \frac{\partial^2}{\partial X_J^2} \Phi_0 
= \sum_{1\leq J<K\leq\cN} \gamma_{J,K}V(X_J-X_K) +2\lambda|\vm|\frac1{\Phi_0}\frac{\partial}{\partial\beta}\Phi_0 -\cE_0
\end{equation} 
with the constant
\begin{equation}
\label{cE0_1} 
\cE_0 = \sum_{1\leq J<K\leq\cN}\Bigl( \lambda^2 m_Jm_K(m_J+m_K)c_0 + 2|\vm|\lambda^2m_Jm_K c_1\Bigr).  
\end{equation} 
This is equivalent to \Ref{Id1} with $\cH$ in \Ref{cH} and $\cE_0$ in \Ref{cE0}. \QED

\subsection{Corollary \ref{cor1}}
\label{sec4.2} 
An outline of how to obtain Corollary~\ref{cor1} from Proposition~\ref{prop1} is given in the main text. For completeness we provide the formal details here. 

We choose $\cN=N+\tilde{N}+M+\tilde{M}$ and set $m_J=1$ for $1\leq J\leq N$, $m_J=-1/\lambda$ for $1\leq J-N\leq \tilde{N}$, $m_J=-1$ for $1\leq J-N-\tilde{N} \leq M$, and $m_J=1/\lambda$ for $1\leq J-N-\tilde{N}-M\leq \tilde{M}$. Denoting $X_J$ as $x_J$ for $1\leq J\leq N$, $\tilde{x}_J$ for $1\leq J-N\leq \tilde{N}$, $y_J$ for $1\leq J-N-\tilde{N} \leq M$, and $\tilde{y}_J$ for $1\leq J-N-\tilde{N}-M\leq \tilde{M}$ we find by straightforward computations that $\cH$ in \Ref{cH} and \Ref{gJK} is equal to $H_{N,\tilde{N}}(\vx,\tilde{\vx})-H_{M,\tilde{M}}(\vy,\tilde{\vy})$ as defined in \Ref{HNtN}, $\Phi_0(\vX)$ in \Ref{Phi0} is proportional to $F_{N,\tilde{N},M,\tilde{M}}(\vx,\tilde{\vx},\vy,\tilde{\vy})$ in \Ref{FNtNMtM}, and $\cE_0$ in \Ref{cE0} is equal to $C_{N,\tilde{N},M,\tilde{M}}$ in \Ref{CNtNMtM}. Since obviously $2\lambda|\vm|=2[\lambda(N-M)-\tilde{N}+\tilde{M}]$ this implies the result. \QED

\subsection{Remark \ref{rem1}} 
\label{sec4.3}
Equations \Ref{cH}, \Ref{Phi0} and \Ref{Id1} imply that, redefining the elliptic functions by $\beta$-dependent constants
\begin{equation}
\begin{split}
V(r)&\to V(r) + b_0\\
\tet(r)&\to B_1\tet(r), \quad b_1:=\frac{\partial}{\partial\beta}\log B_1
\end{split} 
\end{equation} 
changes the constant in \Ref{cE0_1} as follows,
\begin{equation}
\begin{split} 
\cE_0 \to & \cE_0 -\sum_{J<K} \gamma_{J,K} b_0 -  
2\lambda|\vm|\sum_{J<K}\lambda m_J m_K b_1 \\ = & \cE_0 - \Bigl(\lambda^2(|\vm^2||\vm|-|\vm^3|)-\lambda(\cN-1)|\vm|\Bigr)b_0 - \lambda^2(|\vm|^2-|\vm^2|)|\vm|b_1   
\end{split} 
\end{equation}
(we inserted \Ref{gJK} and made a straightforward computation).  The claim made in Remark~\ref{rem1} corresponds to the special case $b_0=c_0$ and $b_1=c_1$ (some further details are explained in \cite{EL4}, Remark~1.1).\QED

\subsection{Lemma  \ref{lem1}} 
\label{sec4.4}
We note that the identity in \Ref{Id1} is invariant under 
\begin{equation}
\label{subs1}
\begin{split} 
\Phi_0(\vX)&\to c \Phi_0(\vX)\ee{\ii v\sum_{J=1}^{\cN}m_JX_J}\\ 
\cE_0&\to \cE_0+ |\vm|v^2
\end{split}
\end{equation}
for real $v$ and non-zero complex $c$. Indeed, the function $\Phi_0(\vX)$ in \Ref{Phi0} is obviously invariant under translations $X_J\to X_J+a$, for all $a\in\R$, and this implies
\begin{equation}
\sum_{J=1}^{\cN}\frac{\partial}{\partial X_J}\Phi_0(\vX)=0  
\end{equation} 
and
\begin{equation}
\cH \Phi_0(\vX)\ee{\ii v\sum_{J=1}^{\cN}m_JX_J} = 
\ee{\ii v\sum_{J=1}^{\cN}m_JX_J}\Bigl( \cH +|\vm|v^2\Bigr)\Phi_0(\vX). 
\end{equation} 
Using the generalization of \Ref{Id1} obtained by the substitutions in \Ref{subs1} and restricting to the special case as in Section~\ref{sec4.2} we obtain the result in Lemma~\ref{lem1}. \QED 

The substitution in \Ref{subs1} has a natural physical interpretation as a change of the center-of-mass velocity of the particle system. 

\bigskip 

\noindent \textbf{Note added:} The integrability of the CFSV type deformation of the eCS model in \Ref{HNtN} for $\mathcal{N}=1$ and $V(r)$ in \Ref{V} was proved in \cite{K}.\footnote{I thank an anonymous referee for pointing this out to me.}

\bigskip 

 \noindent \textbf{Acknowledgments.}  I am grateful to Martin Halln\"as for helpful comments. This work was supported by the Swedish Science Research Council (VR) and the G\"oran Gustafsson Foundation.

\appendix 

\section{Elliptic functions}
\label{appA}

For the convenience of the reader we collect here various properties of the elliptic functions $V(r)$ in \Ref{V} and $\tet(r)$ in \Ref{tet} that we use in Section~\ref{sec2}.

The function
\begin{equation}
\label{phi} 
\phi(r) := \frac{\partial}{\partial r}\log(\tet(r)) 
\end{equation}
obeys the relations
\begin{equation}
\label{p} 
\frac{\partial}{\partial r}\phi(r) = -V(r) 
\end{equation} 
and 
\begin{equation}
\label{pp}
\phi(r)^2 = V(r) -2f(r) -c_0
\end{equation} 
with the function 
\begin{equation}
\label{f} 
f(r) : = -\frac{\partial}{\partial\beta}\log(\tet(r)) + c_1
\end{equation} 
and the constants $c_{0,1}$ in \Ref{c0c1}. Moreover,
\begin{equation}
\label{ppp} 
\phi(x)\phi(y)+\phi(x)\phi(z) +\phi(y)\phi(z) = f(x)+f(y)+f(z)\; \mbox{ if }\; x+y+z=0 
\end{equation} 
with the same function $f$ in \Ref{f}. We also use  
\begin{equation} 
\label{evenodd} 
\phi(-r)=-\phi(r),\quad V(-r)=V(r),\quad f(-r)=f(r). 
\end{equation} 
All these identities are classic; see \cite{EL2} and \cite{EL4} for references and elementary proofs.

\end{document}